# New interaction potentials for borate glasses with mixed network formers

Siddharth Sundararaman[a,b], Liping Huang[a,†], Simona Ispas[b] and Walter Kob[b]

[a]Department of Materials Science and Engineering, Rensselaer Polytechnic Institute, Troy, NY 12180, United States

[b]Laboratoire Charles Coulomb (L2C), University of Montpellier, CNRS, F34095 Montpellier, France

**Abstract**

We adapt and apply a recently developed optimization scheme used to obtain effective potentials for aluminosilicate glasses to include the network former boron into the interaction parameter set. As input data for the optimization, we used the radial distribution functions of the liquid at high temperature generated by *ab initio* molecular dynamics simulations, and density, coordination and elastic modulus of glass at room temperature from experiments. The new interaction potentials are shown to reproduce reliably the structure, coordination and mechanical properties over a wide range of compositions for binary alkali borates. Furthermore, the transferability of these new interaction parameters allows mixing to reliably reproduce properties of various boroaluminate and borosilicate glasses.

## 1. Introduction

Borate glasses, especially borosilicate glasses, are widely used in a variety of applications ranging from laboratory glassware to optical fibers to nuclear waste glasses, and are among the most scientifically interesting materials today[1–6]. The multiple coordination environments that boron can adopt depending on composition and thermal/pressure history[7–12] have made it much more challenging to simulate using classical simulations as compared to silicate glasses. Attempts to use classical molecular dynamics (MD) simulations to provide insights into structure-property correlations in these materials have resulted in a long history of potential development[13–22].

Some of the earlier successful potentials for these systems include the one by Park and Cormack[14] that has boron coordination number dependent parameters or by Huang and Kieffer[15] that includes charge transfer and three-body interactions. Inoue et al.[17] proposed a potential based on the Born-Mayer-Huggins functional form that includes anionic charges dependent on composition. This potential is able to predict the coordination trends well, but a major deficiency of this potential is that it is reliable only when the samples are quenched in the NVT (constant number of atoms, constant volume and constant temperature) ensemble. Large discrepancies are observed in both magnitude and trends for coordination and density when quenched in the NPT (constant number of atoms, constant pressure and constant temperature) ensemble.

---

†Corresponding author. E-mail address: huangL5@rpi.edu (L. Huang).

For borosilicate glasses, Kieu et. al.[16] proposed a pairwise potential that includes composition dependent parameters. This potential is able to reproduce structural and mechanical properties of the sodium borosilicate system fairly well. Later this potential was extended to include aluminum for boroaluminosilicate glasses by Deng and Du[18]. A major issue with this potential is that it has both charge and boron energy parameters dependent on the composition and fitted to match experimental results well, so when compositions outside the fitting range were simulated the results were not reliable[21]. Wang et al.[21] optimized parameters based on the Guillot-Sator[23] potential and showed significant improvement over Kieu's potential[16] for soda-lime borosilicate glasses[21].

Recently, Deng et al.[22] extended the parameters proposed by Teter[24–26] to borate systems. The basic idea was very similar to that of Kieu's potential[16], i.e., to have a composition dependent boron energy parameter that was fit specifically to the experimental trends but with constant partial charges. The reduction in the number of fitting parameters significantly improved the properties over various compositions in the sodium borosilicate system considered and showed excellent trends for density and coordination[22]. As discussed above with Kieu's potential[16], this potential has a similar issue due to empirical fitting to coordination trends with system specific parameters and hence is not easily transferrable to systems other than borosilicate glasses.

In our previous work[27], we used an optimization scheme similar to the one developed for silica[28] and extended the interaction parameter set to include alkali modifiers lithium, sodium and potassium, alkaline earth modifier calcium, and aluminum that can behave as a modifier or a former depending on the composition[29,30]. In this work, we adapt a similar optimization approach to include the network former boron into our parameter set to allow MD simulations of glasses with mixed network formers. One of the major goals of our potential optimization scheme is hence to not have any system specific parameters to ensure easy transferability and extensibility to complex multi-component systems.

The organization of the paper is as follows: First, in Sec. 2, the details of how the optimization scheme was adapted to include the network former boron are discussed. Then, in Sec. 3, the new potentials are compared with both *ab-initio* data and experimental data over a large range of compositions and demonstrate their reliability and transferability. Finally, in Sec. 4, the results are summarized and conclusions are drawn.

## 2. Simulations Methods
In this section, we show how the optimization scheme developed by us in our previous work[28] was adapted to include the network former boron into the set of potentials for glasses with mixed network formers.

### 2.1 Potential and cost function
Similar to our previous work on silica glass and aluminosilicate glasses[27,28], we use the Buckingham potential functional form[31] for short-range interactions and the Wolf truncation method[32,33] to evaluate the Coulombic interactions.

$$V^{Buck}(r_{\alpha\beta}) = A_{\alpha\beta}exp(-B_{\alpha\beta}r_{\alpha\beta}) - \frac{C_{\alpha\beta}}{r_{\alpha\beta}^6} + \frac{D_{\alpha\beta}}{r_{\alpha\beta}^{24}} + V^W(r_{\alpha\beta}) \quad (1)$$

where

$$V^W(r_{\alpha\beta}) = q_\alpha q_\beta \left[\frac{1}{r_{\alpha\beta}} - \frac{1}{r_{cut}^W} + \frac{(r_{\alpha\beta} - r_{cut}^W)}{(r_{cut}^W)^2}\right] \quad (2)$$

and $\alpha, \beta \in \{O, Si, Li, Na, K, Al, B\}$. All the parameters from our previous work[27,28] were maintained constant and oxygen charge was evaluated for each compositions in order to maintain charge neutrality by using a similar scheme as described previously[27,34]. The short-range interactions were cut off at 8 Å while the Coulombic interactions were cut off at 10 Å. The simulations were carried out using the LAMMPS (Large-scale Atomic/Molecular Massively Parallel Simulator)[35] software package with a timestep of 1.6 fs. A smaller timestep of 0.8 fs was used during the optimization process when exploring the parameter space to avoid large temporary forces that may arise.

The cost function for optimizing the parameters is similar to the one in our previous work[27,28] and is given by

$$\chi^2(\phi) = w_1 \int_0^{r_{N_{RDF}}} \sum_{\alpha,\beta} (rg_{\alpha\beta}^{calc,3000K}(r|\phi) - rg_{\alpha\beta}^{ref,3000K}(r))^2 dr$$
$$+ w_2(E^{calc,300K}(\phi) - E^{ref,300K})^2 + w_3(\rho^{calc,300K}(\phi) - \rho^{ref,300K})^2 \quad (3)$$
$$+ w_4(C^{calc,300K}(\phi) - C^{ref,300K})^2$$

where $\phi$ is the current parameter set, $\alpha, \beta$ are the different species, $w_1, w_2, w_3, w_4$ are the weights for each contribution, $rg_{\alpha\beta}(r)$ is the radial distribution function (RDF) weighted by the distance $r$ up to a maximum distance of $r_{N_{RDF}} = 7$ Å at 3000 K, $\rho$ is the density, $E$ is the Young's modulus and $C$ is the average boron coordination at 300 K and 0 GPa pressure. The superscript "ref" refers to the first principles or experimental reference data towards which the optimization was carried out, and superscript "calc" refers to the calculated properties using the current parameter set. The major difference in the cost function from our previous work on aluminosilicates[27] is the inclusion of the average boron coordination. It is very important to note here that unlike previous work in literature[16,18,22] we do not include the experimental trends in the composition in our cost function or have system specific or composition dependent parameters to reproduce experimental results. Instead, just the value of coordination at a single composition is used to give the optimizer an idea of the numerical value. The Levenberg-Marquardt[36,37] algorithm is used to minimize the cost function for the optimization.

The RDFs for "calc" were calculated by equilibrating a sample of 1200 atoms for pure $B_2O_3$ and 1380 atoms for alkali borate at 3000 K at a density about 15% less than the corresponding glass density (see Table 3) for given composition for 30 ps in the NVT ensemble, followed by a production run of 40 ps. The reduced density was used to avoid high pressures in the reference data for faster diffusion during the equilibration of liquid at high temperature. For the various binary borate glasses, we used $0.2X_2O–0.8B_2O_3$ ( $X \in Li, Na, K$ ) as the reference systems at high temperature. Due to the lack of satisfactory potentials that can produce reliable room temperature samples for all of the alkali borates, the procedure discussed in our previous work[27] to produce samples at room temperature needed for the optimization was slightly modified as described below.

1. As a first step, an optimization that included only the high temperature structure in the cost function ( $w_2, w_3, w_4 = 0$ ) was used to produce an initial potential. The so-obtained potential was then used to quench liquids to 300 K at zero pressure in the NPT ensemble to generate an initial glass structure.
2. These samples were then used in an optimization that included both the high temperature structure of liquid and density of glass in the cost function ( $w_2, w_4 = 0$ ) to produce another potential which was then used to quench liquids to 300 K at zero pressure in the NPT ensemble to generate an improved glass structure with a more accurate glass density.
3. The samples so generated were then used in subsequent optimizations that further included the Young's modulus and coordination in the cost function.

The method described above is justified based on a previous observation that mechanical properties have a strong dependence on the interaction potential and different potentials can predict very similar structures even if the predicted elastic moduli differ significantly[38]. Even though there is no change in the topology of the structure during the optimization since we don't re-melt, the improved interaction parameters result in improved properties as was shown in our previous work on aluminosilciates[27] and will be shown in this paper. It is also important to note here that the iterative process we described above already produces samples that have decent coordination and since we do not re-melt the samples, changes in the average coordination tend to be small during the optimization. The coordination here serves like an added feedback mechanism by penalizing drastic changes in the interaction parameters to improve a property like the Young's modulus at the expense of the coordination. Furthermore, these samples were used only during the optimization process in order to save computation time and all results shown in this paper are from new samples quenched using the final parameter set shown in Table 1 and 2.

The density and average boron coordination at room temperature were calculated during the optimization by relaxing the quenched samples in the NPT ensemble at 300 K and zero pressure with the current parameter set. The Young's modulus was then calculated by compressing and expanding the samples at 300 K along one direction at a constant strain rate (1.25 ns$^{-1}$) up to a linear change of 0.6% and measuring their stress response:

$$E_x = \frac{d\sigma_x}{d\varepsilon_x} \quad (4)$$

where $E_x$, $\sigma_x$ and $\varepsilon_x$ are the Young's modulus, stress and strain, respectively, along the $x$ direction.

The optimized partial charges and short-range interaction parameters for the different systems are given in Table 1 and Table 2, respectively. The new potentials will be referred to as "SHIK" (Sundararaman, Huang, Ispas, Kob) in the rest of the paper.

**Table 1:** Charge for different species.

| Species | Si | Li | Na | K | Al | B |
|---|---|---|---|---|---|---|
| Charge (e) | 1.7755 | 0.5727 | 0.6018 | 0.6849 | 1.6334 | 1.6126 |

**Table 2:** Short-range interaction parameters.

| i-j | $A_{ij}$ (eV) | $B_{ij}$ (Å$^{-1}$) | $C_{ij}$ (eV·Å$^6$) | $D_{ij}$ (eV·Å$^{24}$) |
|---|---|---|---|---|
| O-O | 1120.5 | 2.8927 | 26.132 | 16800 |
| O-Si | 23108 | 5.0979 | 139.70 | 66.0 |
| Si-Si | 2798.0 | 4.4073 | 0.0 | 3423204 |
| O-Li | 6745.2 | 4.9120 | 41.221 | 70.0 |
| Si-Li | 17284 | 4.3848 | 0.0 | 16800 |
| Li-Li | 2323.8 | 3.9129 | 0.0 | 3240 |
| O-Na | 1127566 | 6.8986 | 40.562 | 16800 |
| Si-Na | 495653 | 5.4151 | 0.0 | 16800 |
| Na-Na | 1476.9 | 3.4075 | 0.0 | 16800 |
| O-K | 258160 | 5.1698 | 130.77 | 16800 |
| Si-K | 268967 | 4.3289 | 0.0 | 16800 |
| K-K | 3648.0 | 4.4207 | 0.0 | 16800 |
| O-Al | 21740 | 5.3054 | 65.815 | 66.0 |
| Al-Al | 1799.1 | 3.6778 | 100.0 | 16800 |
| O-B | 16182 | 5.6069 | 59.203 | 32.0 |
| B-B | 1805.5 | 3.8228 | 69.174 | 6000.0 |
| Li-B | 4148.6 | 3.5726 | 102.36 | 16800 |
| Na-B | 3148.5 | 3.6183 | 34.000 | 16800 |
| K-B | 1548.6 | 2.7283 | 201.36 | 16800 |
| B-Si | 4798.0 | 3.6703 | 207.00 | 16800 |

It is important to note that similar to our previous work on aluminosilicates[27], it was observed that there was no requirement for Al-B short-range interactions. This is probably due to the intermediate behavior of aluminum where it can behave as both a network former and modifier depending on the composition[29,30].

## 2.2 Generation of reference data

The reference data required for structure at high temperatures is from *ab initio* MD simulations performed using the Vienna ab initio package (VASP)[39,40]. The Kohn–Sham (KS) formulation of the density functional theory[41] with generalized gradient approximation (GGA) and the PBEsol (modified Perdew-Burke-Ernzerhof) functional[42,43] was used to describe the electronic structure. The projector-augmented-wave formalism[44,45] was used for the electron-ion interaction for Kohn-Sham orbitals expanded in the plane wave basis set at the Γ point of the supercell with energies up to 600 eV. The electronic convergence criterion for the residual minimization method-direct inversion in iterative space[40] was fixed at $5\times10^{-7}$ eV. These parameters were chosen based on previous studies performed on pure silica glass[28] and sodium borosilicate glasses[46].

*Ab initio* MD simulations were carried out in the NVT ensemble at 3000 K using the Nosé thermostat[47] to control the temperature and by starting from configurations obtained from equilibrium classical MD simulations at the same temperature. A cubic system of N atoms with periodic boundary conditions was used with the simulation box length fixed to a value corresponding to a density about 15% less than experimental glass density at ambient conditions for each composition[48] (see details in Table 3). A lower density was chosen in order to reduce the pressure for faster diffusion during the equilibration of liquid at high temperature. The simulation for a given composition was stopped once the mean squared displacement (MSD) of the slowest element, i.e., boron, reached ~10 $Å^2$ (about 10 ps), which was sufficient for other species to reach the diffusive regime too. We discard the first 1 to 2.5 ps of the trajectory in each case and use the remaining data for calculating the RDFs.

**Table 3:** Number of atoms and density used to equilibrate the liquid at high temperatures in *ab initio* MD simulations.

| System | N (atoms) | $\rho$ (g/cm$^3$) |
|---|---|---|
| $B_2O_3$ | 400 | 1.56 |
| $0.2Li_2O–0.8B_2O_3$ | 460 | 1.79 |
| $0.2Na_2O–0.8B_2O_3$ | 460 | 1.89 |
| $0.2K_2O–0.8B_2O_3$ | 460 | 1.81 |

## 2.3 Generation of samples

Glasses of various compositions, as shown in Table 4 were prepared using the melt-quench method. Samples with ~10000 atoms were first equilibrated in the NVT ensemble for about 100 ps at about the experimental glass density for the composition at a temperature $T_i$ (~1.5 times the simulated glass transition temperature of the composition for the parameter set) and then in the NPT ensemble for about 700 ps at 0.1 GPa. They were then subsequently quenched to 300 K in the NPT ensemble at a nominal quench rate of ~1 K/ps. The small pressure of 0.1 GPa was applied at temperature $T_i$ as a precaution against the system entering the gas phase at high temperature, which was ramped down to 0 GPa during the quenching process. The samples were then annealed at 300 K and 0 GPa for 100 ps in the NPT ensemble. Four independent samples were quenched for each composition to improve the statistics of the results.

**Table 4:** Details of the quenching process for each glass system.

| Composition | $T_i$ (K) |
|---|---|
| $xY_2O$–$(1-x)B_2O_3$ ($Y \in$ Li, Na, K; $x \in 0, 0.05, 0.1, 0.15, 0.2, 0.25, 0.3, 0.35, 0.4, 0.45, 0.5$ ) | 1850 |
| $RNa_2O$–$B_2O_3$–$1SiO_2$ ( $R \in 0.1, 0.3, 0.5, 0.7, 0.9, 1.1, 1.3, 1.6$ ) | 1400-2100 |
| $RNa_2O$–$B_2O_3$–$2SiO_2$ ( $R \in 0.1, 0.3, 0.5, 0.7, 0.9, 1.1, 1.3, 2, 3, 4$ ) | 1400-2300 |
| $RNa_2O$–$B_2O_3$–$3SiO_2$ ( $R \in 0.5, 0.75, 1, 1.25, 1.5, 2, 2.5, 3, 4, 5$ ) | 1400-1800 |
| $0.25Na_2O$–$xAl_2O_3$–$(0.75-x)B_2O_3$ ( $x \in 0.05, 0.1, 0.15, 0.2, 0.25, 0.3$ ) | 2500 |

## 3. Results and Discussion

In this section, the ability of these new potential parameters to reproduce properties both in the liquid and the glassy state of various boron containing systems will be discussed. In Sec 3.1, we will demonstrate the ability of the new potentials to predict various properties for alkali borates both in the liquid state and the glassy state. We will then show in Sec 3.2, that these new parameters can be used in conjunction with parameters from our previous study on aluminosilicates to simulate ternary aluminoborate glasses. Finally, in Sec 3.3 we will look at the ability of our potential to predict various properties and trends for borosilicate glasses.

### 3.1 Pure $B_2O_3$ and alkali borates
### 3.1.1 Structure of liquids

We first compare the equilibrium structure of the liquid as predicted by these new potentials with the one obtained from *ab initio* MD simulations. Figure 1(a) shows the partial RDFs predicted by these potentials for pure $B_2O_3$ at 3000 K and Fig. 1(b-d) show some partial RDFs for alkali borate systems compared to data from *ab initio* simulations.

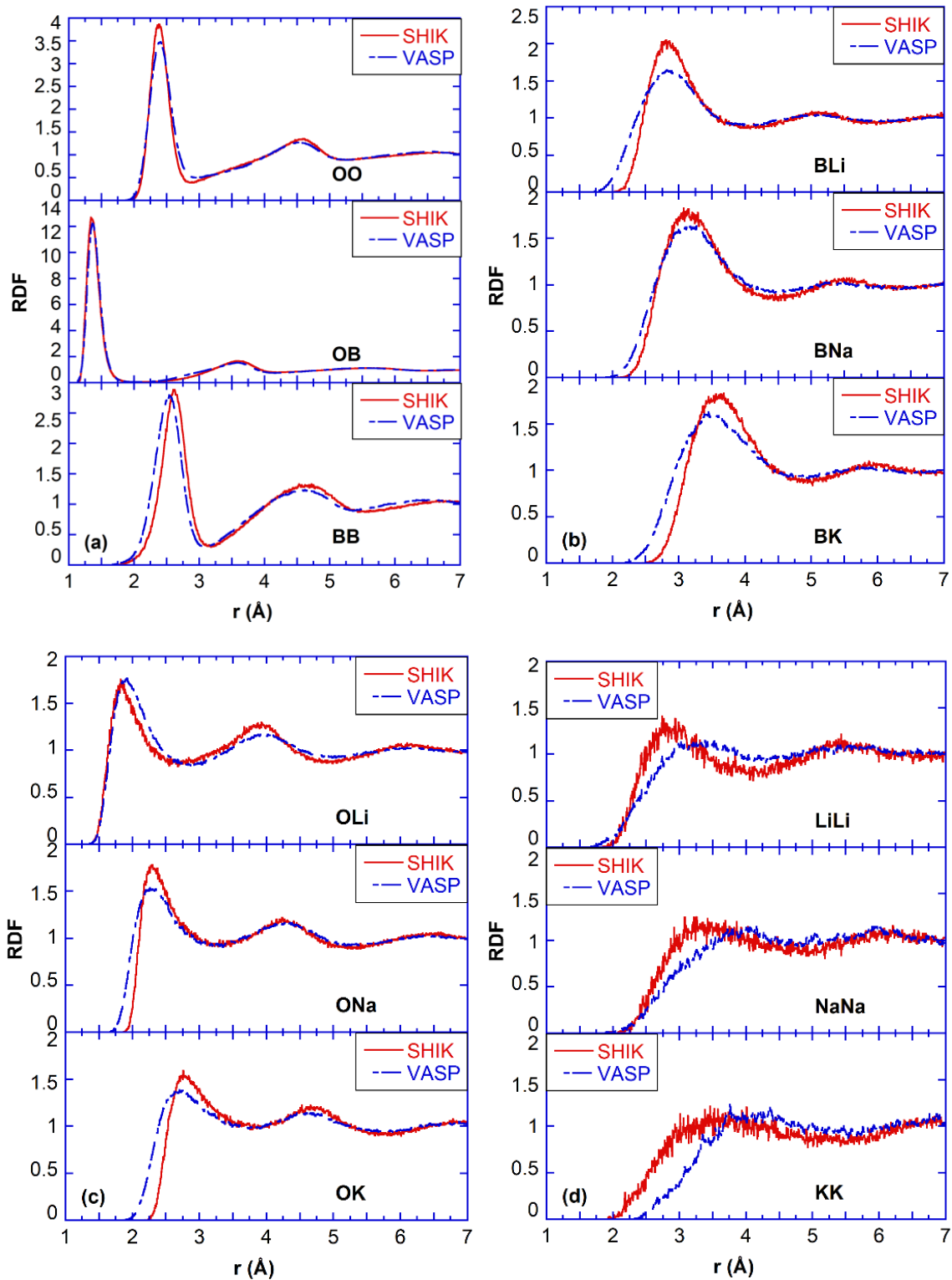

**Fig. 1** Partial radial distribution functions as obtained from the SHIK potential (red solid line) and *ab initio* simulations (blue dashed line) at 3000 K. (a) O-O, O-B and B-B pairs in pure $B_2O_3$, (b) B-X, (c) O-X and (d) X-X pairs in $0.2X_2O$–$0.8B_2O_3$ where X is Li, Na and K.

Overall the various partial RDFs are predicted very well as compared to the *ab initio* data which as noted before is not surprising since the RDFs are included in the cost function that is minimized. Even though some differences are observed in the intensities of the first peaks in certain RDFs, overall the location of the first peak is very well reproduced. The major deviation from this trend is in the like cation partial RDFs as seen in the B-B partial in the lower panel of Fig. 1(a) and X-X partials in Fig. 1(d) (X is Li, Na, K), where it is observed that for the network former, the first peak is shifted slightly to the right as compared to the *ab initio* data, while for the network modifiers the peaks are shifted slightly to the left resulting in a compensation effect for the density. In fact, for like modifier cations, the *ab initio* structure shows almost no correlation. It is important to note though that there are other quantities in the cost function that are being minimized, like the coordination and density of the glass, and hence a compromise to improve the prediction of these properties could have resulted in slightly deteriorated RDFs for the liquid in classical MD. Overall, we observe that the classical MD structures are more ordered as compared to their *ab initio* counterparts.

### 3.1.2 Density and coordination of alkali borate glasses

Borate glasses are scientifically very interesting materials due to the possibility of different boron coordination environments depending on composition and thermodynamic conditions[7–12] (boron anomaly). With increasing alkali content in alkali borate glasses, initially an increase in the average boron coordination is observed due to the formation of 4-fold coordinated boron, which eventually reaches a maximum at a certain alkali content, and then the average B coordination starts to decrease due to the formation of 3-fold coordinated boron with non-bridging oxygens[7,9,10].

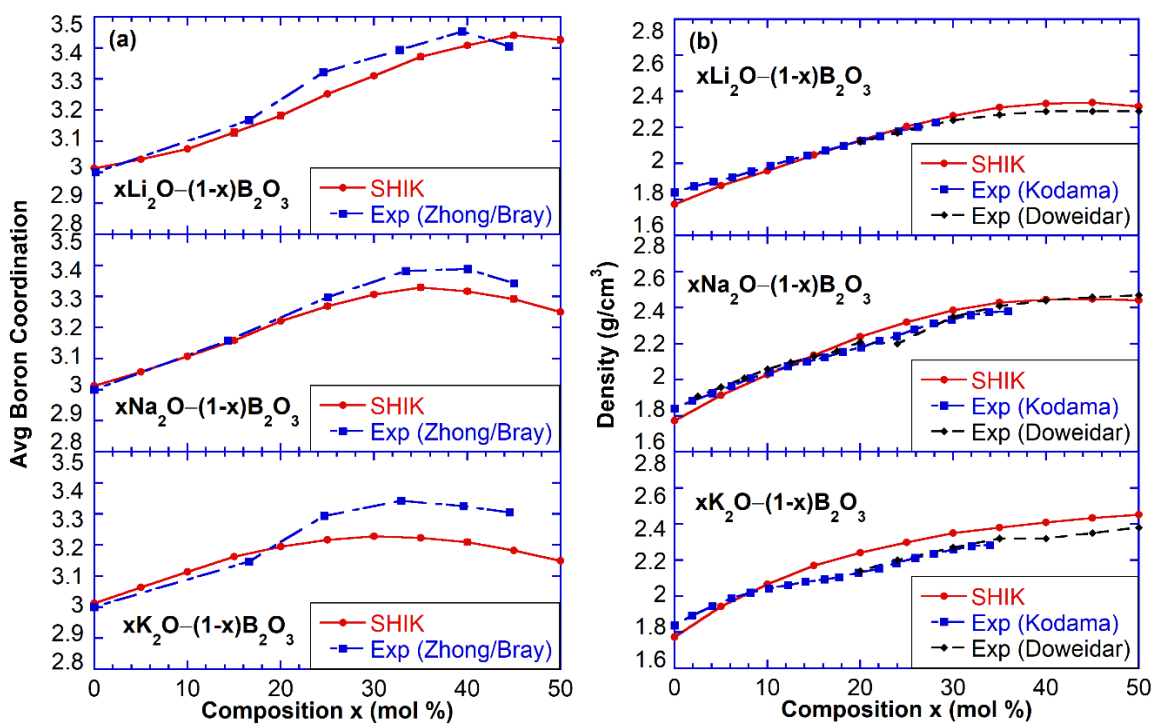

**Fig. 2** (a) Average boron coordination and (b) density as obtained from the SHIK potential (red solid line) compared to experiments[10,48–50] (blue and black dashed line) as a function of alkali content for binary borate systems.

Figure 2(a) shows the average boron coordination predicted by the SHIK potential as compared to experiments[10]. Not just the trends in the coordination as a function of alkali content, but also the numerical values are reproduced very well by the SHIK potential and are within a few percent error. We note here that the numerical values of the coordination of boron in the potassium borate system are lower than experimental values when potassium content is higher than 20%. Attempts to increase the coordination of boron result in further increase in the elastic moduli for these samples which are already relatively high (see Fig. 4), so we chose this compromise. It should further be noted that this discrepancy could also be partially attributed to the presence of water in potassium borate glasses with high potassium content in experiments due to their high affinity for water[51]. For the case of sodium and potassium, the maximum is predicted almost perfectly while for lithium it is predicted at a slightly higher alkali content. It is important to remember here that these new potentials do not have any empirical fitting to the coordination trends and no pre-described parameters dependent on these trends, which are different from the other potentials in the literature for borates[16,18,22]. We believe that these trends come out naturally largely due to the charge balancing scheme that is an effective way to capture the partial ionic-covalent nature of these systems.

One of the simplest quantities to compare with experiments is the density of glass. Figure 2(b) shows the variation of the glass density for the $xY_2O–(1-x)B_2O_3$ (Y is Li, Na and K) systems as a function of modifier content[48–50]. These densities are predicted very well as compared to experiments. At compositions close and above the corresponding maximum of the boron coordination, the density slightly decreases in lithium borate, plateaus in sodium borate and slightly increases in potassium borate with alkali content, respectively. This is due to the reduction of 4-fold coordinated boron thereafter compensated by the addition of alkali atoms with increasing atomic weight.

To further understand why the so-called boron anomaly occurs, Fig. 3 shows the calculated fraction of B atoms in different coordination environments in $xNa_2O–(1-x)B_2O_3$ system. $^3B_{BO}$ and $^3B_{NBO}$ refers to 3-fold coordinated boron with only bridging oxygen and with at least one non-bridging oxygen, respectively, their sum being $^3B_{tot}$ which is the fraction of total number of 3-fold coordinated boron, and finally $^4B_{total}$ refers to the fraction of total number of 4-fold coordinated boron. As sodium initially is added to $B_2O_3$, it mainly converts 3-fold coordinated boron to 4-fold coordinated boron. At higher sodium concentrations the fraction of 3-fold coordinated boron with non-bridging oxygen increases very quickly. This is where the total number of 3-fold coordinated boron starts to increase and that of the 4-fold coordinate boron starts to decrease, resulting in the maximum in the average boron coordination in Fig. 2(b) which is in agreement with the accepted Dell, Bray, Xiao (DBX) model[52] for such systems.

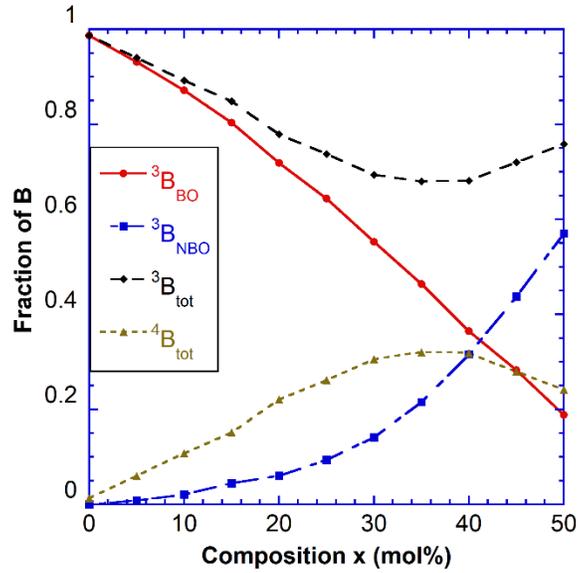

**Fig. 3** Fraction of B atoms in different coordination environments as a function of Na$_2$O content in xNa$_2$O–(1-x)B$_2$O$_3$.

### 3.1.3 Elastic moduli of alkali borate glasses

Elastic moduli of glasses are of great practical interest. Figure 4(a-b) show the Young's modulus and the bulk modulus predicted by these potentials as compared to experimental values[48]. Once again, we observe that for the alkali borate glasses, the trends in the elastic moduli are predicted very well by the SHIK potential. The numerical values for the elastic moduli at high alkali content are overestimated especially for the bulk modulus, and as we mentioned before in Sec 3.1.2 we chose a compromise between various properties to aim for their trends as a function of composition to be consistent with experimental observations. A plateau or maximum is observed for the Young's modulus at compositions very close to the corresponding maximum of the average boron coordination and can be explained from the reduction of the 4-fold coordinated boron thereafter.

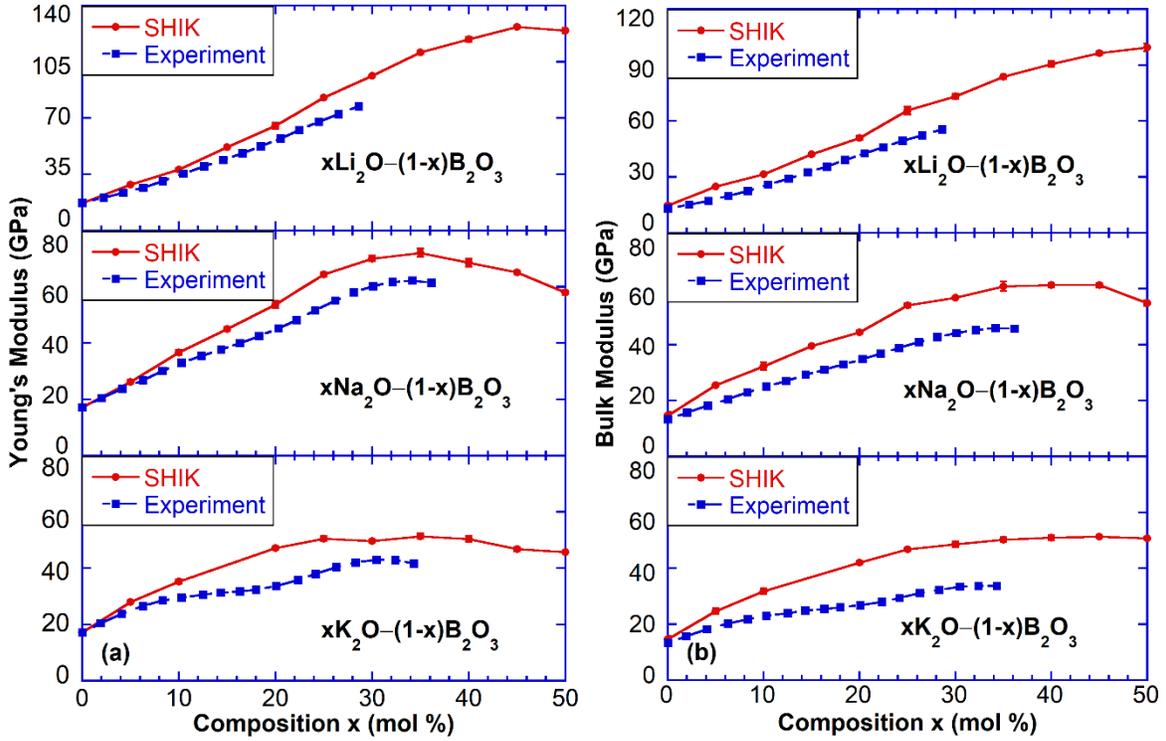

**Fig. 4** (a) Young's modulus and (b) bulk modulus as obtained from the SHIK potential (red solid line) compared to experiments[48] (blue dashed line) as a function of alkali content for alkali borates.

### 3.2 Sodium aluminoborates

Aluminum is another important element in glasses that can play a dual role of both a former and a modifier depending on the composition[29,30]. The average aluminum and boron coordination in sodium aluminoborates predicted by the SHIK potential are plotted as a function of alumina content as compared to experimental data[53] in Fig. 5(a-b). Aluminum tends to behave as a network former and is predominantly in 4-fold coordination state in sodium aluminoborates[54]. Fig. 5(a) shows that the average B coordination decreases with the increase of $Al_2O_3$ content in the system at a constant sodium content[53,54]. This indicates that sodium tends to prefer charge balancing the 4-fold coordinated Al over 4-fold coordinated boron. Higher coordinated (5-fold and 6-fold) Al is observed at high Al content when there is not sufficient Na to charge balance the 4-fold coordinated Al. Compared to experiments[53], both the numerical values and the decreasing trend of the B average coordination and the increasing trend of the Al average coordination with the increase of $Al_2O_3$ content are reproduced well by the SHIK potential.

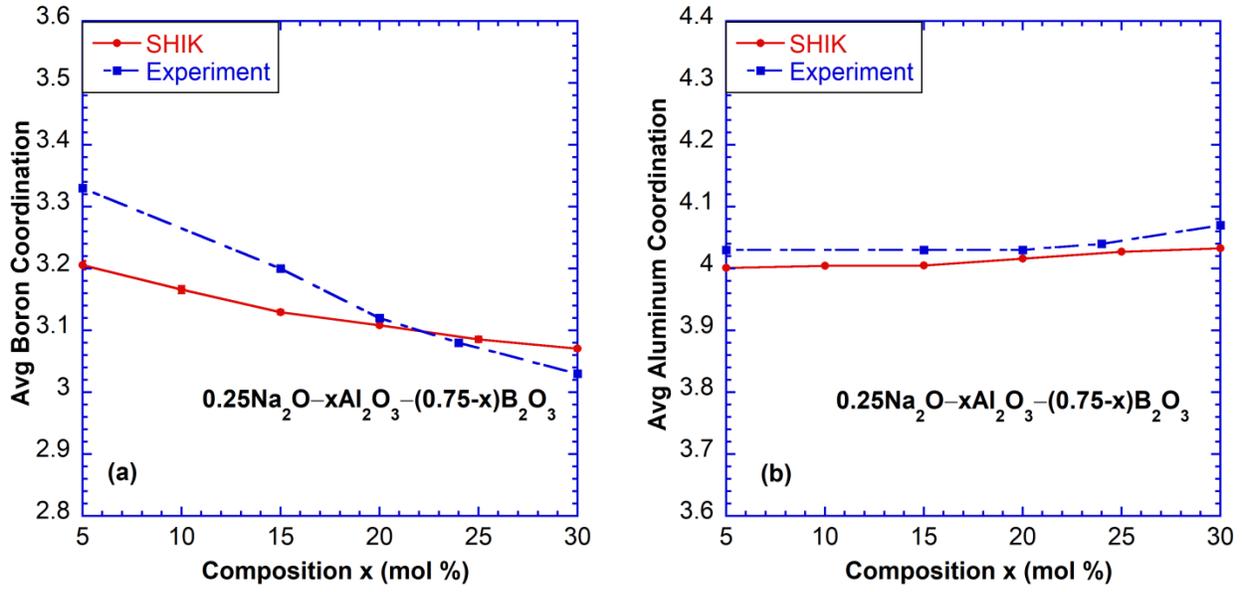

**Fig. 5** Average coordination of (a) B and (b) Al as obtained from the SHIK potential (red solid line) compared to experiments[53] (blue dashed line) as a function of $Al_2O_3$ content for the $0.25Na_2O–xAl_2O_3–(0.75-x)B_2O_3$ system.

Figure 6 shows the density of the $0.25Na_2O–xAl_2O_3–(0.75-x)B_2O_3$ system as a function of alumina as compared to experiments[53]. The aluminum parameters are from our previous work[27] and as mentioned before in Section 2.1, we observed no need for any Al-B interactions. As the $Al_2O_3$ content is increased, the density first decreases and then increases with a shallow minimum between 10 and 20% $Al_2O_3$. This is attributed to the competition between the substitution of the $B_2O_3$ by the heavier $Al_2O_3$ and the decreasing packing efficiency of the glassy network[53]. The increase in density after 20% $Al_2O_3$ can be attributed to the increasing amount of 5 and 6-fold coordinated Al as seen in Fig. 5(b), while the initial decrease in density is due to the decreasing average boron coordination as seen in Fig. 5(a). It can be seen in Fig. 6 that both the trend and the numerical values for the density are reproduced within a few percent error by the SHIK potential as compared to experiments[53].

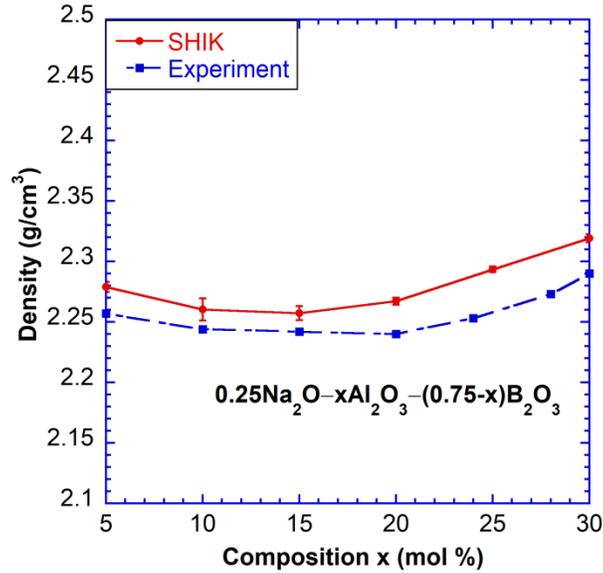

**Fig. 6** Density as obtained from the SHIK potential (red solid line) compared to experiments[53] (blue dashed line) as a function of $Al_2O_3$ content for the $0.25Na_2O–xAl_2O_3–(0.75-x)B_2O_3$ system.

To further validate the SHIK potential, elastic moduli of aluminoborate glasses were calculated for these compositions and show in Fig. 7. We again observe that both the Young's modulus and the bulk modulus, similar to the trends in density, decrease initially due to the decrease in the average boron coordination and increase after ~20% $Al_2O_3$ due to the increase in the amount of 5-fold and 6-fold coordinated Al. Figure 7 shows that the SHIK potential is also able to reliably predict the trends in the elastic properties for the compositions studied here even though the numerical values are overestimated as we have discussed before in Section 3.1.3.

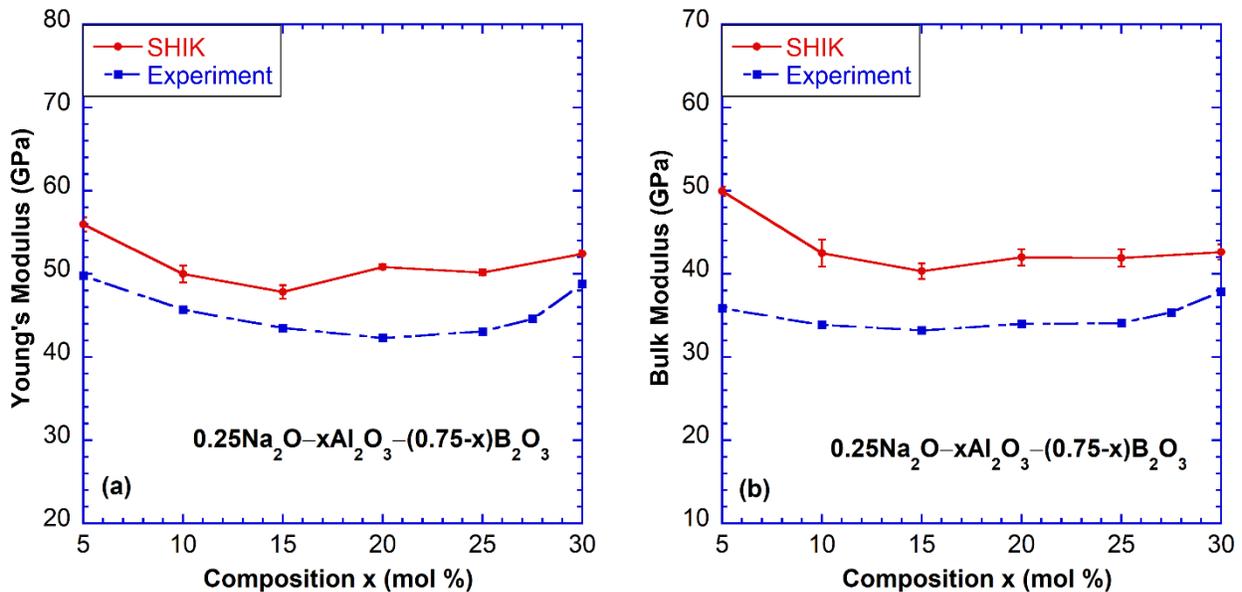

**Fig. 7** (a) Young's modulus and (b) bulk modulus as obtained from the SHIK potential (red solid line) compared to experiments[53] (blue dashed line) as a function of $Al_2O_3$ content for the $0.25Na_2O$–$xAl_2O_3$–$(0.75-x)B_2O_3$ system.

### 3.3 Sodium borosilicates

The final system studied using the new set of interaction parameters is the sodium borosilicate glasses. Figure 8(a-c) show the average boron coordination for the systems $RNa_2O$–$B_2O_3$–$KSiO_2$ where $K = \dfrac{\#SiO_2}{\#B_2O_3}$ for K=1, 2, 3, respectively, with varying R where $R = \dfrac{\#Na_2O}{\#B_2O_3}$. In these systems, the average boron coordination increases initially with sodium converting $^3B$ to $^4B$ which eventually reaches a maximum, then stays constant over a composition range where the sodium creates non-bridging oxygens associated with Si, and then decreases at a higher alkali content where the sodium starts to produce 3-fold coordinated boron with non-bridging oxygens ($^3B_{NBO}$)[8,52]. Our simulations predict these trends in experimental data[8] very well though the maximum in the average boron coordination is predicted at a slightly higher alkali content. Figure 8(d-f) show the density changes for the systems $RNa_2O$–$B_2O_3$–$KSiO_2$ corresponding to those in (a)-(c). We note that the trends in the density where a maximum is reached after a certain alkali content are predicted very well with numerical values within 10% of the experimental data[55,56]. Furthermore, we observe that the plateau is reached at compositions close to the maximum in the average boron coordination for the corresponding systems and hence can be attributed to the reduction of 4-fold coordinated boron and the addition of $Na_2O$ content.

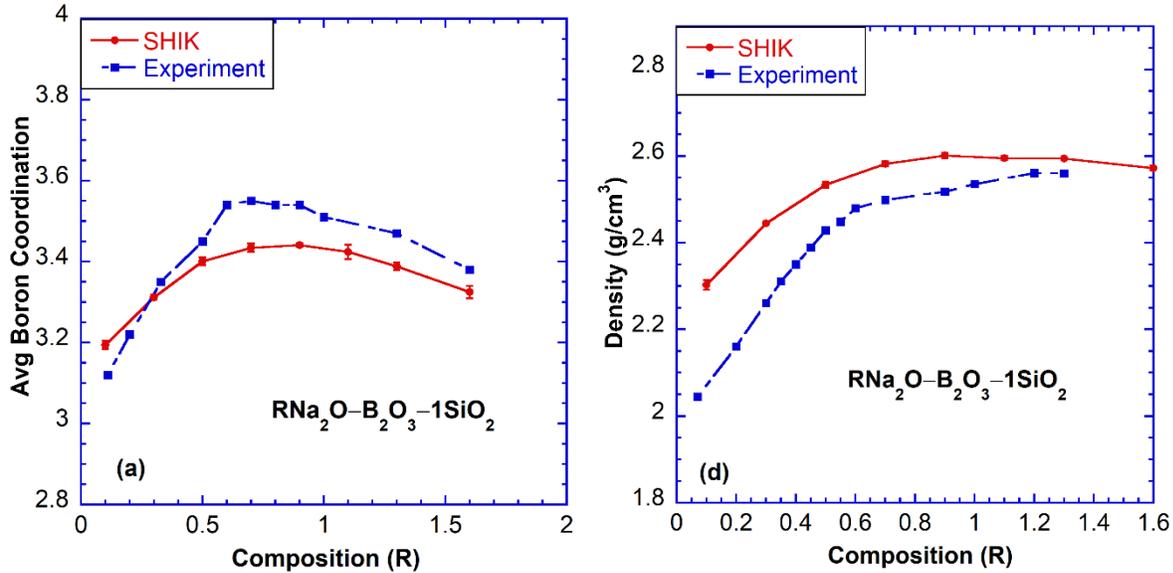

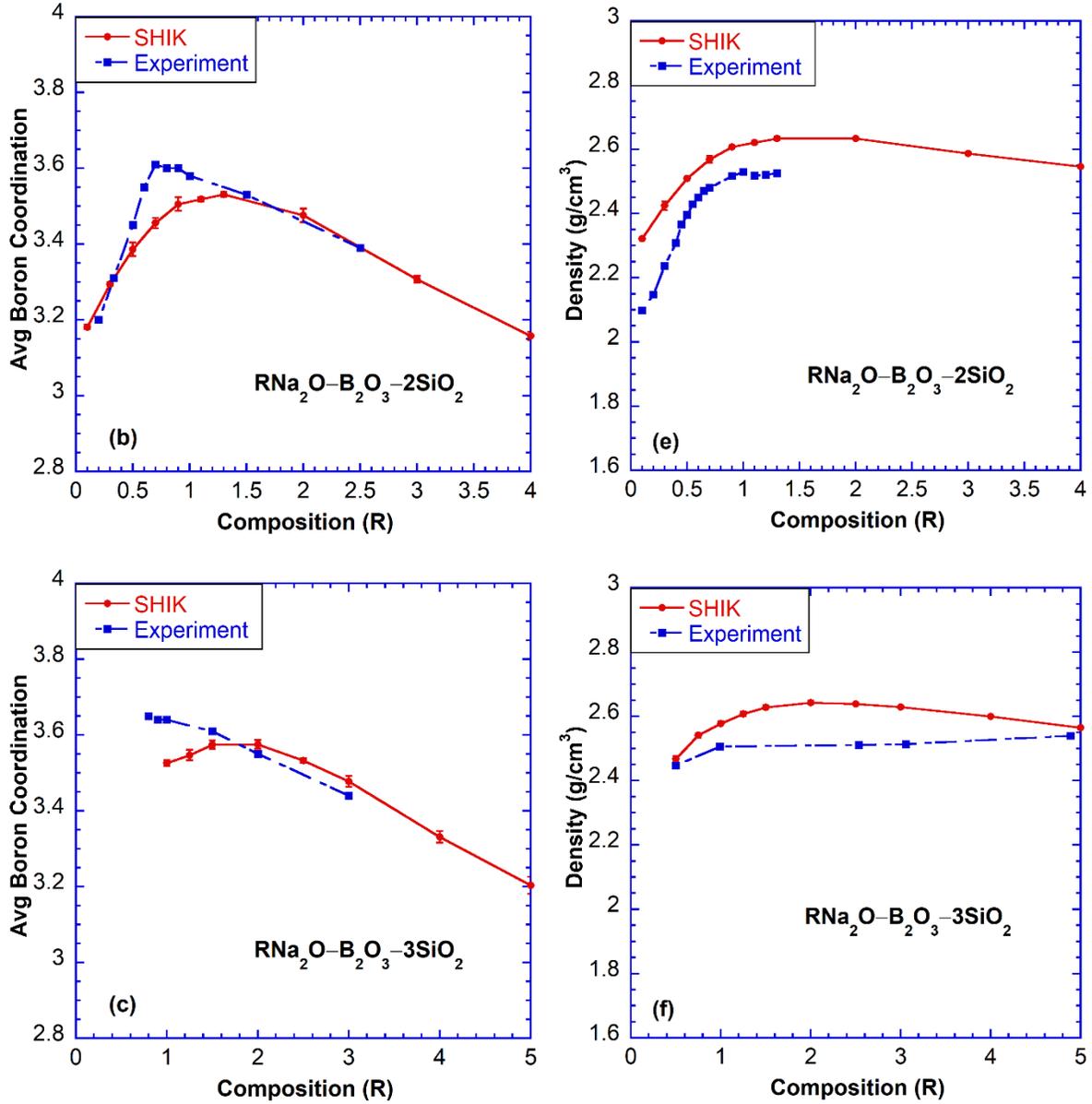

**Fig. 8** (a), (b), (c) Average boron coordination and (d), (e), (f) density as obtained from the SHIK potential (red solid line) compared to experiments[8,55,56] (blue dashed line) as a function of Na$_2$O content for the RNa$_2$O–B$_2$O$_3$–1SiO$_2$, RNa$_2$O–B$_2$O$_3$–2SiO$_2$, RNa$_2$O–B$_2$O$_3$–3SiO$_2$ system, respectively, with varying R where $R = \frac{\# Na_2O}{\# B_2O_3}$.

Figure 9(a)-(b) show the average boron coordination for sodium borosilicate glasses in the RNa$_2$O–B$_2$O$_3$–KSiO$_2$ system (where K is 0, 1, 2 and 3) as a function of the modifier content R from experiments[8] and the SHIK potential. The location and height of the maximum in the average boron coordination depends on the SiO$_2$ to B$_2$O$_3$ ratio (i.e., K), with a higher K resulting in the maximum at a higher average boron coordination and at a higher alkali content. This trend is also predicted very well by the SHIK potentials though the alkali content at which the maximum occurs is higher than in experiments.

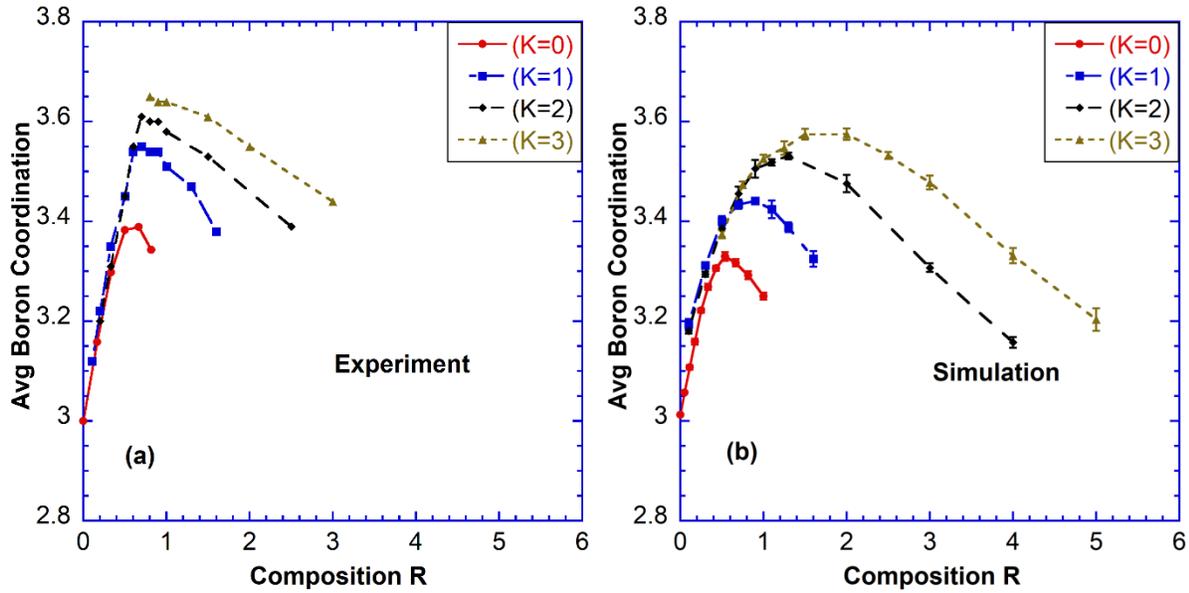

**Fig. 9** Average boron coordination for different series of sodium borosilicates for the RNa$_2$O–B$_2$O$_3$–KSiO$_2$ system, where K is 0, 1, 2 and 3, from (a) experiments[8] and (b) the SHIK potential.

To analyze the structure further, the neutron structure factor of the 3Na$_2$O–B$_2$O$_3$–6SiO$_2$ glass calculated from the SHIK potential is compared to the one from experiments[57] in Fig. 10. The new potential not only reproduces the peak positions from experiments correctly but also their intensities well. This shows that the structure of sodium borosilicate system can be very well described by the SHIK potential.

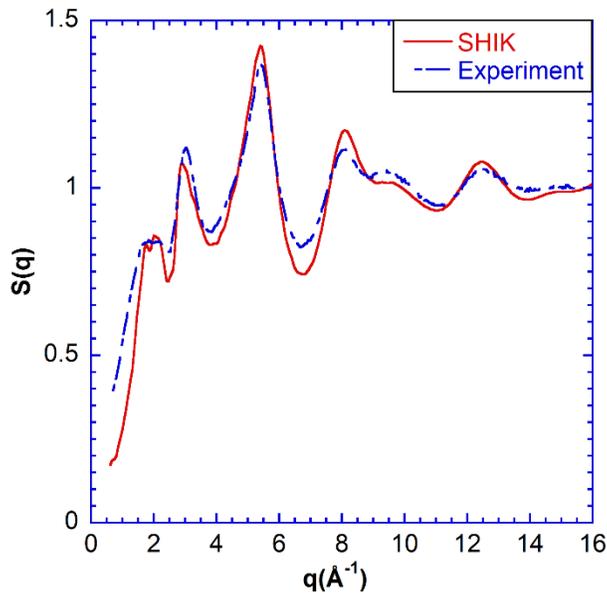

**Fig. 10** Neutron structure factor as obtained from the SHIK potential (red solid line) compared to experiments[57] (blue dashed line) for the $3Na_2O-B_2O_3-6SiO_2$ glass.

Finally, Fig. 11(a-b) show the elastic moduli for the $RNa_2O-B_2O_3-SiO_2$ system as a function of the modifier content R calculated from the SHIK potential as compared to experiments[56]. Trends in elastic moduli are predicted very well with the maximum at the correct alkali content. We observe that for this composition, the bulk modulus plateaus, and the Young's modulus shows a maximum at compositions close to the maximum in boron coordination and can hence also be attributed to the subsequent decrease in the average boron coordination. The numerical values of the elastic moduli are higher than measured from experiments[56] especially at very low sodium content, but over the rest of the composition range studied, the errors are within 15% of experimental values.

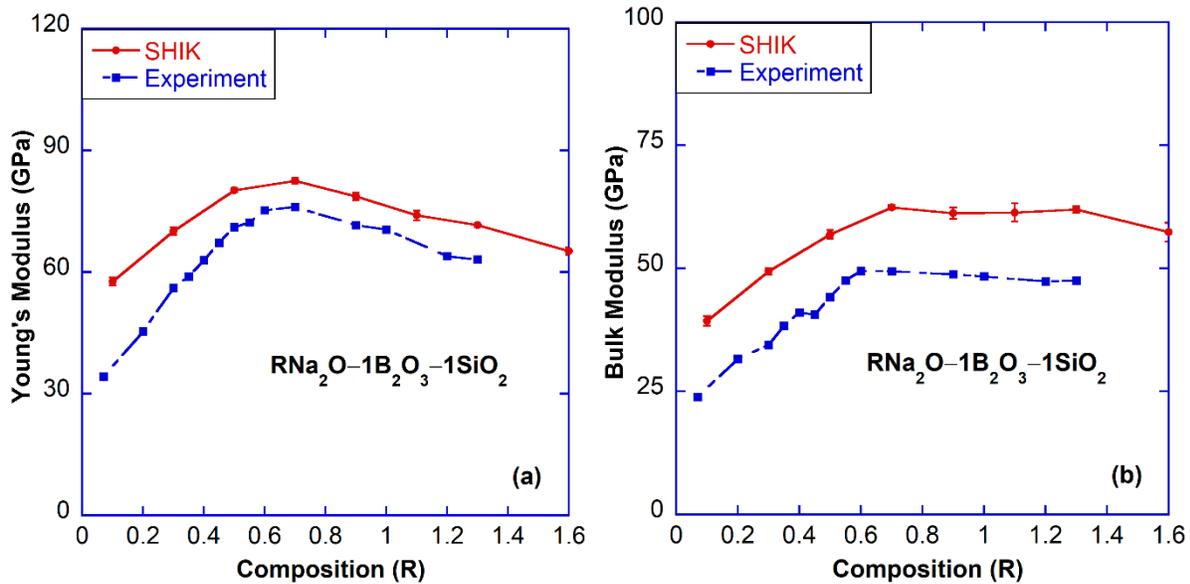

**Fig. 11** (a) Young's modulus and (b) bulk modulus as obtained from the SHIK potential (red solid line) compared to experiments[56] (blue dashed line) as a function of $Na_2O$ content for the $RNa_2O-B_2O_3-SiO_2$ system.

## 4. Summary and Conclusions

In the present work we have used an optimization scheme similar to the one developed earlier[27,28] to include the network former boron into our set of interaction parameters by reproducing the liquid structure at high temperature, and the density, coordination and elastic modulus of the glass at room temperature. Parameters from the previous optimization for silica and aluminosilicates[27,28] were maintained to ensure transferability and a similar charge balancing scheme suggested by Habasaki et al.[34] was used to partially emulate the polarization effect.

The new set of interaction parameters is shown to predict both the trends and numerical values of different properties reliably over a wide range of compositions in glass systems with mixed network formers. Challenges associated with the boron coordination changes with composition and thermodynamic conditions were overcome without any system specific fitting or pre-described parameter dependence. This allows for easy transferability that enables the exploration of structure and properties of new multi-component glasses not yet synthesized in experiments. These reliable pairwise potentials also allow for high computational efficiency to study large and complex systems.

## Acknowledgements

L. Huang acknowledges the financial support from the US National Science Foundation under grant No. DMR-1105238 and DMR-1255378, and a Corning-CFES seed fund, as well computational resources from the Center for Computational Innovations (CCI) at RPI. S. Ispas acknowledges HPC resources from GENCI (Grants A0010907572 and A0030907572). S. Sundararaman acknowledges an NSF-IMI travel grant from DMR-0844014 to France to initiate this collaborative work. W. Kob is a member of IUF.